%% file: LP99.tex
\documentstyle[12pt,epsfig]{article}
\newcommand{\sgrav}{\mbox{$ \widetilde{{\rm G}}$}}
\newcommand{\kizer}{\mbox{$ \widetilde{\chi}^0_{\rm i}$}}
\newcommand{\kione}{\mbox{$ \widetilde{\chi}^0_1$}}
\newcommand{\mkione}{\mbox{$ m_{\widetilde{\chi}^0_1}$}}
\newcommand{\kitwo}{\mbox{$ \widetilde{\chi}^0_2$}}
\newcommand{\kipl}{\mbox{$ \widetilde{\chi}_1^+$}}
\newcommand{\kimi}{\mbox{$ \widetilde{\chi}_1^-$}}
\newcommand{\charg}{\mbox{$ \widetilde{\chi}_1^{\pm}$}}
\newcommand{\mcharg}{\mbox{$ m_{\widetilde{\chi}_1^{\pm}}$}}

\newcommand{\eea}{\mbox{$ {\rm e}^+{\rm e}^- \rightarrow$}}
\newcommand{\pp}{\mbox{$ {\rm p}{\bar {\rm p}} \rightarrow$}}
\newcommand{\sle}{\mbox{$ \widetilde{\rm l}$}}

\newcommand{\selr}{\mbox{$ \widetilde{\rm e}_{\rm R}$}}

\newcommand{\smur}{\mbox{$ \widetilde{\mu}_{\rm R}$}}
\newcommand{\sta}{\mbox{$ \widetilde{\tau}$}}
\newcommand{\staur}{\mbox{$ \widetilde{\tau}_{\rm R}$}}
\newcommand{\stau}{\mbox{$ \widetilde{\tau}_1$}}

\newcommand{\stoone}{\mbox{$ \widetilde{\rm t}_1$}}

\newcommand{\sbotone}{\mbox{$ \widetilde{\rm b}_1$}}
\newcommand{\snu}{\mbox{$ \widetilde{\nu}$}}
\newcommand{\msnu}{\mbox{$ m_{\widetilde{\nu}}$}}
\newcommand{\snue}{\mbox{$ \widetilde{\nu}$}$_{\rm e}$}
\newcommand{\snut}{\mbox{$ \widetilde{\nu}$}$_{\tau}$}
\newcommand{\snum}{\mbox{$ \widetilde{\nu}$}$_{\mu}$}
\newcommand{\sq}{\mbox{$ \widetilde{\rm q}$}}
\newcommand{\sg}{\mbox{$ \widetilde{\rm g}$}}
\newcommand{\Rp}{$\rm{R_p}$\hspace{-1.em}{\rotatebox{-10}{/}}\ }
\newcommand{\Rpar}{$\rm{R_p}$}

\newcommand{\Gevm}{\mbox{GeV/$c^2$}}
\newcommand{\rs}{\mbox{$\sqrt{s}$}}
\newcommand{\tb}{\mbox{$\tan\beta$}}
\newcommand{\mA}{\mbox{$\rm{m_A}$}}
\newcommand{\mh}{\mbox{$\rm{m_h}$}}
\newcommand{\hpm}{\mbox{$\rm{H}^{\pm}$}}
\newcommand{\hpl}{\mbox{$\rm{H}^{+}$}}

\input LP99macros.tex

\def\Title#1{\begin{center} {\Large {\bf #1} } \end{center}}

\begin{document}

\Title{New particle searches}

\bigskip\bigskip


\begin{raggedright}  

{\it V. Ruhlmann-Kleider \index{Ruhlmann-Kleider, V.}\\
DSM/DAPNIA/SPP, Saclay,\\
91191 Gif-sur-Yvette Cedex, FRANCE }
\bigskip\bigskip
\end{raggedright}

\section{Introduction}

  This review covers a few selected topics from the searches performed at
Tevatron, HERA and LEP2. Details on the data samples analysed at the time 
of the conference are given in Table~\ref{tab:data}. 

\begin{table}[htb]
\begin{center}
\begin{tabular}{|c|c|c|c|c|c|} \hline  
collider & experiments & beams &  \rs &
            period & $\L$/expt \\
         & & & & & (pb$^{-1}$) \\ \hline
Tevatron & CDF/D0 & {\bf p}\mbox{\boldmath $\bar{\rm{p}}$} &
           1.8~TeV &'87/'96  & $\sim$ 110 \\

HERA & H1/ZEUS & {\bf e$^+$p} & 300~GeV & '94/'97  & $\sim$ 40 \\
     &         & {\bf e$^-$p} & 318~GeV & '98/'99  & $\sim$ 15 \\

LEP2 & ALEPH/DELPHI & {\bf e$^+$e$^-$} 
     &  130 to 183~GeV & '95/'97 & $\sim$ 90 \\         
     & L3/OPAL &{\bf e$^+$e$^-$} &189~GeV & '98 &   $\sim$ 170 \\
     & &{\bf e$^+$e$^-$} &192/196~GeV & '99 & $\sim$ 105 \\ \hline

\end{tabular}
\caption{Data samples analysed for the '99 summer conferences by
experiments at Tevatron, HERA and LEP2. The last column gives the 
integrated luminosity per experiment for each of the data taking periods.}
\label{tab:data}
\end{center}
\end{table}

New particle searches are dominated by LEP whose results come mostly 
from data up to 189~GeV except for some updates from '99 data up to 196~GeV.
Combined results from the LEP experiments also exist on some subjects and
are denoted ADLO.

After a brief outline of the searches on exotic particles, results
on supersymmetric particles and Higgs bosons are detailed.
All exclusion limits are at the 95\% Confidence Level.

\section{Exotic particles}

  Searches for exotic particles encompass a great variety of topics, 
such as technicolor particles~\cite{technicolor}, 
new Z' bosons~\cite{zprime,lepindirect}, 
four fermion contact interactions~\cite{contact,lepindirect}, 
new, excited or exotic fermions~\cite{fstar}, 
leptoquarks. Constraints are derived at the three colliders either from direct
searches or from comparing precise measurements with 
Standard Model (SM) expectations.

 
  As an illustration, the case of leptoquark (LQ) searches is detailed. The
phenomenology of leptoquarks is describbed by three parameters: 
M$_{\rm LQ}$, the LQ mass, $\lambda_{\rm{lq}}$ and $\beta_{\rm l}$,
the LQ coupling and branching ratio into a given pair of SM
lepton and quark, (l,q). Results are interpreted either assuming leptoquarks
to be coupled to a single SM generation, with fixed $\beta_{\rm l}$ (as in the
BRW model~\cite{brw}) or in more generic models, with $\beta_{\rm l}$ 
variable and possible mixed couplings.

\begin{table}[htb]
\begin{center}
\begin{tabular}{|c|c|c|c|c|} \hline  
collider & LQ type & assumptions &  limit (\Gevm ) & couplings \\ \hline
Tevatron & 1st gen. LQ & $\beta_{\rm e}$=1 & 
           242 & any $\lambda_{\rm{eq}}$ \\
Tevatron & 2nd gen. LQ & $\beta_{\mu}$=1 & 
           202 & any $\lambda_{\mu\rm{q}}$ \\
Tevatron & 3rd gen. LQ & $\beta_{{\nu}_{\tau}}$=1 & 
           149 & any $\lambda_{{\nu}_{\tau}\rm{q}}$ \\
Tevatron & 3rd gen. LQ & $\beta_{\tau}$=1 & 
            99 & any $\lambda_{\tau\rm{q}}$ \\
HERA     &  1st gen. LQ & BRW model & 
           265 & $\lambda_{\rm eq} = 0.3$\\ 
LEP2     & 1st or 2nd gen. S$_{1,L}$ & BRW model &
           370 & $\lambda_{\rm lq} = 0.3$\\
\hline
\end{tabular}
\caption{Limits on leptoquarks from Tevatron, HERA and LEP2. Specific values
have been assumed for the LQ branching ratios and/or couplings to SM particles.
Precise measurement at LEP2 allow to probe high masses in some specific
cases only.}
\label{tab:lq}
\end{center}
\end{table}

\begin{figure}[htb]
\begin{center}
\epsfig{file=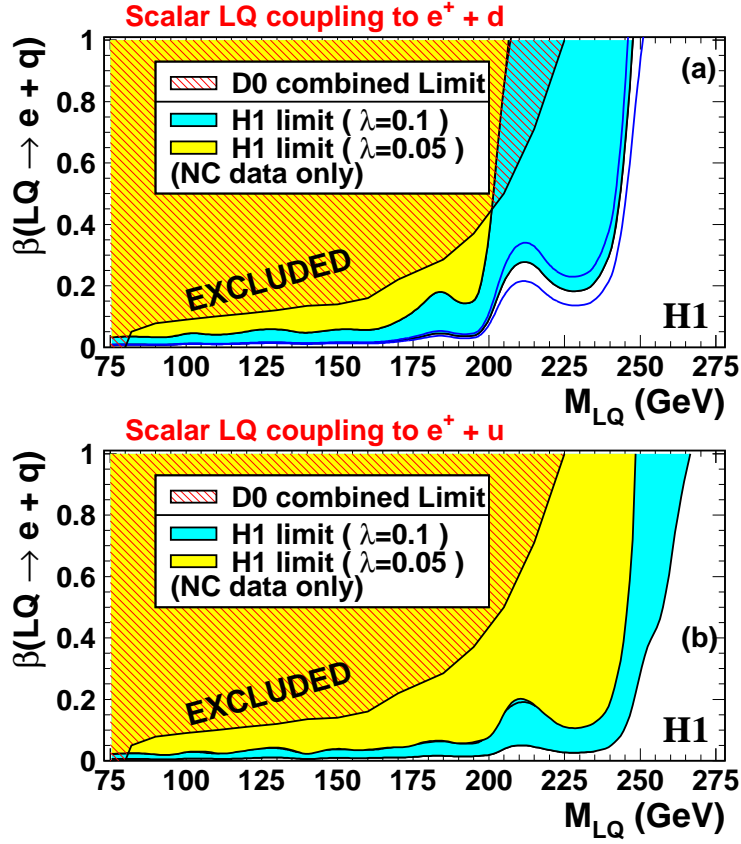,width=11cm}
\caption{Constraints on first generation leptoquarks from H1 and D{\O}.
For $\lambda = 0.1$, the error bands illustrate the sensitivity to d and u 
quark densities.}
\label{fig:lq}
\end{center}
\end{figure}

  The current constraints in the first approach~\cite{lqh1,lq} 
are summarized in Table~\ref{tab:lq}. 
The interpretation in generic models has been pioneered
by H1~\cite{lqh1}. As an example, Figure~\ref{fig:lq} shows the regions
excluded by H1 and D{\O} in the plane M$_{\rm LQ}$ vs $\beta_{\rm e}$ for 
first generation leptoquarks. HERA sensitivity extends down to small
$\beta_{\rm e}$ even for small couplings, \eg\ for $\beta_{\rm e} \sim 10\%$
and $\lambda_{\rm eq} \sim 0.05$ leptoquark masses up to 200~\Gevm\ are
excluded. The interesting case of mixed couplings has also been 
studied~\cite{lqh1}. To cite one result, leptoquarks with couplings to
both (e, q) and ($\tau$, q) have been found to be excluded with a
sensitivity similar to that quoted for pure first generation leptoquarks.

\section{Supersymmetric particles}
All supersymmetric particle searches are conducted within the
Minimal Supersymmetric Standard Model (MSSM) with additional assumptions
to decrease the number of free parameters. Depending on those assumptions,
the phenomenology differs and so do the experimental signatures.
At present, three theoretical frameworks are studied.

\subsection{Constrained MSSM}\label{sec:cmssm}
The most common framework 
assumes R-parity conservation and soft supersymmetry breaking mediated 
by gravity. Soft breaking terms are thus unified at high energy 
(the so-called GUT scale) and the number of free parameters is reduced 
to five: the common 
sfermion\footnote{Tevatron experiments use a somewhat different and
more constrained scheme, called minimal supergravity (mSUGRA), 
which defines $\rm{m_0}$ as a common scalar mass term at the 
GUT scale and assumes in addition radiative EW symmetry breaking, 
so that \mbox{$\mu$} is fixed up to a sign.}
mass term at GUT scale, $\rm{m_0}$, 
the common gaugino mass term at GUT scale, $\rm{m}_{1/2}$, 
the common trilinear coupling at GUT scale, A$_0$,
the  Higgs mixing parameter, $\mu$,
and the ratio of the two Higgs doublet vacuum expectation values, \tb.

The phenomenology at low energy is derived using renormalisation group
equations. The lightest supersymmetric particle (LSP) is in most cases the 
lightest neutralino, \kione. Due to R-parity conservation, sparticles
are produced in pairs and decay in their SM partner and a sparticle. At the
end of the decay chain, the LSP appears and, as it is stable, gives rise to
missing energy. Results on all types of sparticles have been reported at
the conference.

\subsubsection{Sfermions}
Charged sleptons and light squarks are searched for at LEP 
through the decays summarized in Table~\ref{tab:sfermions}, which are
the dominant ones if the researched sparticle is the next lightest sparticle.
The experimental sensitivity depends on both the sfermion mass and the mass
difference $\Delta M$ between the sfermion and the LSP.
The experimental sensitivity starts from  $\Delta M$ 
above a few \Gevm\ and covers sfermion masses up to 70 to 90~\Gevm\ depending
on the sfermion, as can be seen in Table~\ref{tab:sfermions}
for the LEP combination at 189~GeV~\cite{swg189}. 
Similar sensitivities are reached by the 
individual LEP experiments at 196~GeV~\cite{susy196}. All LEP results
are derived for minimal production cross-sections and hence have a 
general validity.

\begin{table}[htb]
\begin{center}
\begin{tabular}{|c|c|c|c|c|c|} \hline  
experiment & sfermion &decay & obs. limit& exp. limit & $\Delta M$ \\
           &          &      & (\Gevm)   & (\Gevm)    & (\Gevm)  \\ \hline
ADLO    & \selr  & e      \kione & 89 & 90 & $> 15$ \\
189 GeV & \smur  & $\mu$ \kione & 84 & 83 & $> 15$ \\
        & \staur & $\tau$ \kione & 71 & 77 & $> 15$ \\ \hline
ADLO    & \stoone & c \kione      & 87 & 84 & $> 10$ \\ 
189 GeV & \stoone & b l \snu      & 90 & 87 & $> 10$ \\
        & \sbotone& b \kione      & 80 & 68 & $> 10$ \\ \hline
CDF     & \stoone & c \kione      & 89 & -  & $> 40$ \\
run I   & \sbotone& b \kione      & 105& -  & $> 40$ \\ 
\hline
\end{tabular}
\caption{Lower limits on sfermion masses. Combined LEP results up to 189~GeV
are given as well as CDF results from run I.
The LEP expected limits are computed from simulation only, assuming
no signal.
The last column gives the range of validity of the limits expressed as 
a minimal difference between the masses of the sfermion and the LSP
(\kione\ or \snu).}
\label{tab:sfermions}
\end{center}
\end{table}

Light squarks are also searched for at Tevatron~\cite{stopcdf}. 
The experimental sensitivity is complementary from that of LEP since 
it covers higher squark masses and starts at higher $\Delta M$, as 
illustrated in Table~\ref{tab:sfermions}.
Searches for heavy squarks and gluinos belong to Tevatron~\cite{squarkd0}. 
The final states result from \sq\ and \sg\ cascade decays to the LSP and 
quarks, gluons, W 
or Z bosons. The present experimental reach is around 250~\Gevm\ but it must 
be noted that most results are derived for specific values of some 
mSUGRA parameters (A$_0 = 0$, $\mu < 0$ and \tb = 2) which restricts their 
range of validity.

\subsubsection{Charginos and neutralinos}
Due to its excellent coverage of the various signatures of supersymmetry,
LEP provides limits on the masses of the lightest chargino and 
neutralino which are practically absolute
in the constrained MSSM framework.

Direct searches for the lightest chargino \charg\ 
provide limits on \mcharg\
close to the kinematical limit in most of the parameter 
space~\cite{susy196}.
As \kione\ cannot be detected, direct neutralinos searches rely on the
production of heavier neutralinos (\eg\ \kione\ \kitwo\ or \kitwo\ \kitwo\ ) 
and thus bring little constraint on \mkione\ except when \tb\ is close to 1. 
But, combining the results from \kizer\ and \charg\ searches provides 
a limit on \mkione\ valid for large m$_0$ whatever the other 
parameters~\cite{susy196}.

\begin{figure}[htb]
\vspace{-1cm}
\begin{center}
\begin{tabular}{cc}
\epsfig{file=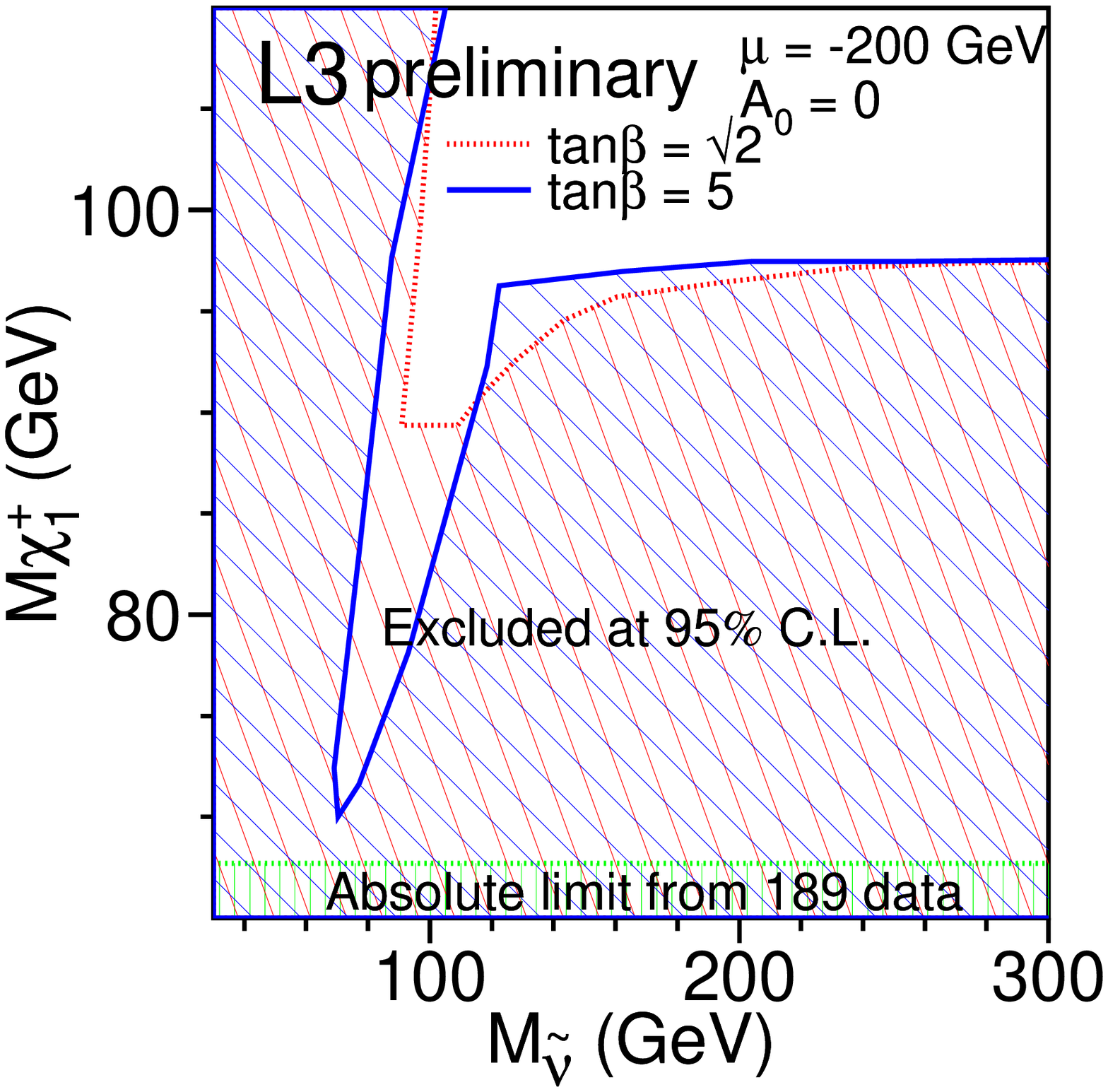,width=8cm} 
&
\epsfig{file=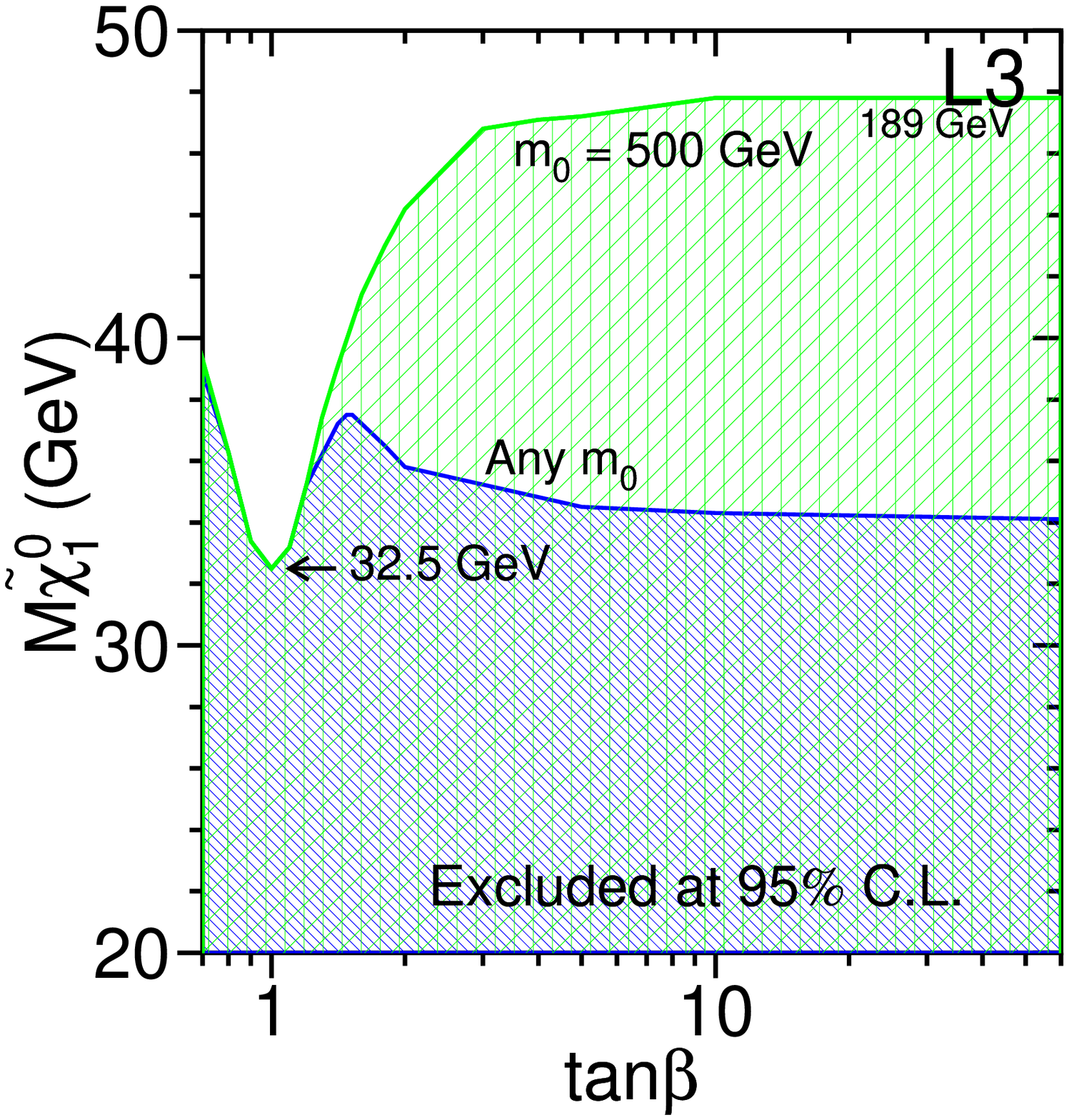,width=8cm} 
\end{tabular}
\caption{Left: L3 Limit on \mcharg\ as a function of \msnu.
The absolute limit from data up to 189~GeV is compared with the '99 limits 
obtained
for two sets of the MSSM parameters.
For a heavy \snu\ the kinematical limit is almost reached
while for a \snu\ around 70~\Gevm, the limits drop.
Right: L3 limit on \mkione\ as a function of
\tb\ from data up to 189~GeV. 
The limit obtained at large m$_0$ whatever the other
parameters is compared with the limit obtained when m$_0$ is also
allowed to vary. The absolute limit corresponds to \tb=1.}
\label{fig:lsp}
\end{center}
\end{figure}

For low values of m$_0$ (that is light \snu), the \charg\ production 
cross-section
drops due to the negative interference between the s-channel production process
and the t-channel \snu\ exchange diagram. The \charg\ decay into
l \snu\ becomes dominant and escapes detection when \charg\ and \snu\
are denegerated. Under the same conditions, the \kizer\ production increases
(because of a positive interference term) but the \kizer\ decay into
$\nu$ \snu\ opens, leading to invisible final states. But, low values of
m$_0$ also mean light sleptons, which are thus within experimental reach.
Combining \kizer, \charg\ and \sle\ searches, limits on \mcharg\ and 
\mkione\ valid for A$_0$=0 and for all values of the other parameters 
can be derived~\cite{lsp189,lspa0}, as illustrated in 
Table~\ref{tab:charg} and Figure~\ref{fig:lsp}. Preliminary studies 
varying A$_0$ (that is allowing for \sta\ mixing)
show that the limit on \mcharg\ may be affected but not that 
on \kione \cite{lspa0}.

\begin{table}[htb]
\vspace{-0.3cm}
\begin{center}
\begin{tabular}{|c|c|c|c|c|} \hline  
experiment & \rs  & searches & limit  & validity \\ 
           & (GeV)& combined & (\Gevm)& \\ \hline
OPAL  & 196 & \charg\ & \mcharg\ $>$ 97.6 
      & large m$_0$, $\Delta M > 10$~\Gevm\ \\
ALEPH & 196 & \charg, \kizer\ & \mkione\ $>$ 34.0 
      & large m$_0$ \\
L3    & 189 & \charg, \kizer, \sle\ & \mcharg\ $>$ 67.7
      & A$_0$=0\\
L3    & 189 & \charg, \kizer, \sle\ & \mkione\ $>$ 32.5
      & absolute \\
\hline
\end{tabular}
\caption{Lower limits on \mcharg\ and \mkione, from LEP.
As more results are used, the range of validity of the limits extends
.}
\label{tab:charg}
\end{center}
\end{table}

\subsection{R-parity breaking}
  In the second theoretical framework, soft supersymmetry breaking is
still mediated by gravity but R-parity (\Rpar) is assumed to be broken via an 
extra-term to the superpotential of the form:

\beq
 \W =  \lambda_{ijk} {\rm L}_i {\rm L}_j \bar{\rm E}_k  +
 \lambda_{ijk}^{'} {\rm L}_i {\rm Q}_j \bar{\rm D}_k  + 
 \lambda_{ijk}^{''} \bar{\rm U}_i \bar{\rm D}_j \bar{\rm D}_k
\eeq{eq:rp}

where $ijk$ denote generation indices and
the capital letters refer to superfields associated to 
left-handed doublets of leptons (L) and quarks (Q), and right-handed 
singlets of charged leptons (E), down-type quarks (D) and up-type quarks (U). 
$\W$ implies violation of the leptonic and baryonic numbers.
In addition to the five parameters related to supersymmetry breaking
($\rm{m_0}$, $\rm{m}_{1/2}$, A$_0$, $\mu$ and \tb), \Rpar\ breaking (\Rp)
introduces 45 couplings (9 $\lambda_{ijk}$, 27 $\lambda_{ijk}^{'}$
and 9 $\lambda_{ijk}^{''}$). For sake of simplicity, searches
are conducted assuming only one coupling to dominate at a time and all
sparticles to decay close to the interaction vertex. This latter hypothesis
corresponds to assuming the \Rp\ couplings to be greater than 
values which are at least two orders of magnitude below the current 
experimental limits on most couplings.

\begin{figure}[htb]
\begin{center}
\begin{tabular}{c|c}
\epsfig{file=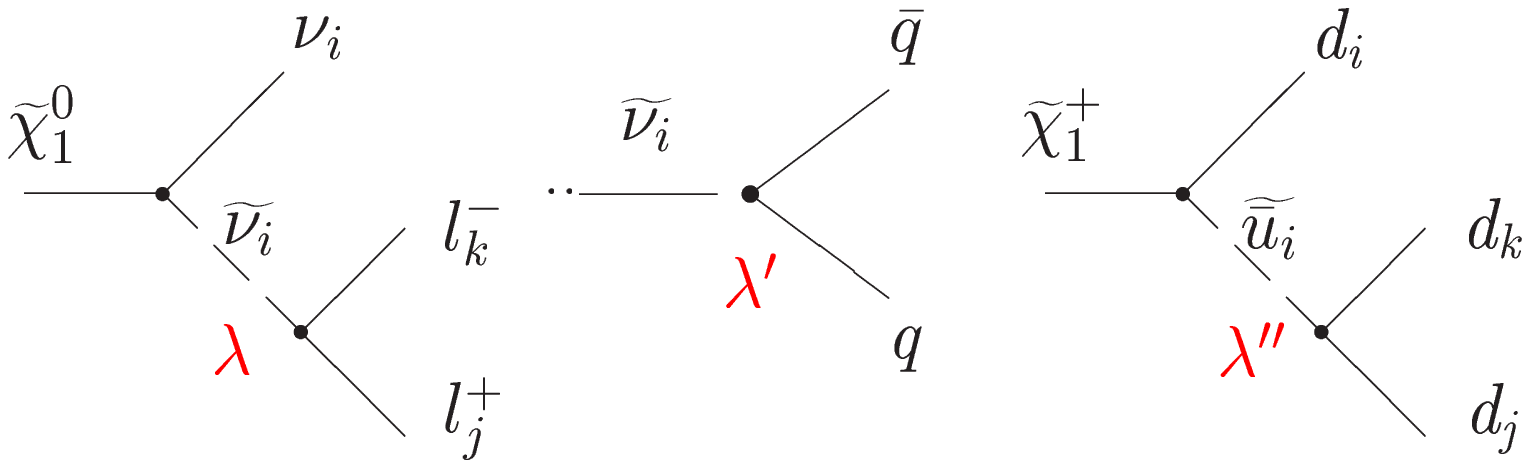,height=2.4cm} &
\epsfig{file=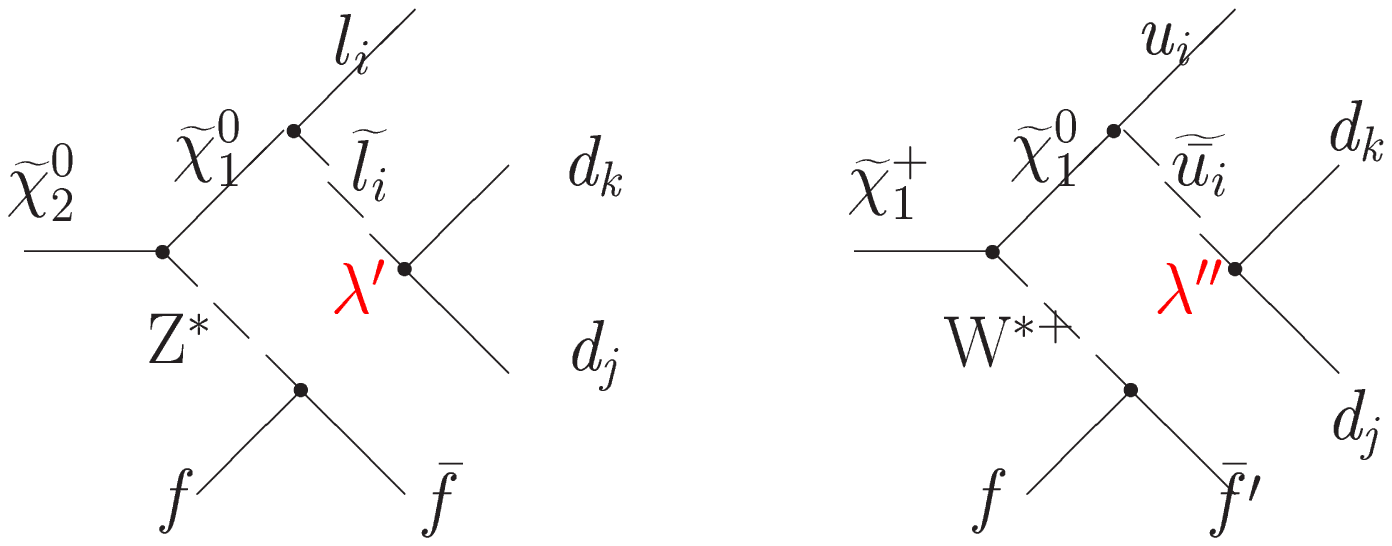,height=2.6cm} \\
a) direct decays & b) indirect decays \\
\end{tabular}
\caption{Examples of decays allowed by \Rpar\ violation.}
\label{fig:diag}
\end{center}
\end{figure}

  Compared to \Rpar\ conservation, \Rp\ modifies the 
phenomenology at low energy. Single sparticle production is possible, the
LSP (which is still \kione\ in most cases) is no longer stable and the 
sparticle decay patterns change a lot. Sparticle can decay into
SM particles through one \Rp\ vertex (\eg\ \snu\ case in
Figure~\ref{fig:diag}a) or via an \Rpar\ conserving vertex leading to
an off-shell sparticle that decays through an \Rp\ vertex
(\eg\ \kione\ and \charg\ cases in Figure~\ref{fig:diag}a). These decays are
referred to as direct decays. A second type of decays, called indirect
decays, implies cascade decays to SM particles through several \Rpar\
conserving and \Rp\ vertices with some sparticles on-shell,
such as in Figure~\ref{fig:diag}b.

  Thus, there are many final states to consider, with 
multileptons and/or multijets and possibly missing energy from neutrinos.
As in the constrained MSSM case, all types of sparticles have been searched
for.

\subsubsection{Charginos and neutralinos}
  As in the \Rpar\ conserving scheme, LEP sets constraints on charginos
and neutralinos which are valid over wide ranges of the underlying 
parameters (except A$_0$ which is set to 0). The \charg\ and \kizer\
production modes are as in the \Rpar\ conserving case while their decay
pattern is completely modified. The most notable change concerns the
LSP whose production leads to observable final states whatever the 
dominant \Rp\ coupling. All experimental signatures expected
from the production and decays of \kipl\ \kimi, \kione\ \kione\ and
\kitwo\ \kione\ have been investigated at LEP2. Due to \kione\ being
detectable, combining \kizer\ and \charg\ searches alone suffices to
derive constraints~\cite{rplep} 
on the sparticle masses valid for  A$_0 = 0$ and 
all values of the other parameters, as illustrated in 
Table~\ref{tab:rpar}.

\begin{table}[htb]
\begin{center}
\begin{tabular}{|c|c|c|c|}\hline
  \Rp\ coupling & \multicolumn{3}{|c|}{limits (\Gevm)} \\
  type & M$_{\widetilde{\chi}_1^0}$ &
   M$_{\widetilde{\chi}_2^0}$ &
   M$_{\widetilde{\chi}_1^{\pm}}$ \\ \hline
  $\lambda,\lambda^{'}$ couplings:
  & 30 & 50 & 94 \\ \hline
  $\lambda^{''}$ couplings:
  & 32 & 67 & 94 \\\hline
\end{tabular}
\caption[]{L3 limits on \kione, \kitwo\ and \charg\ masses in the \Rp\
scheme, from data up to 189~GeV.}
\label{tab:rpar}
\end{center}
\end{table}

\subsubsection{Sneutrinos}
  
  Another appreciable change due to \Rp\ is the allowed decay of sneutrinos
which make them directly observable \eg\ at LEP through double and single
productions.

\begin{figure}[htb]
\begin{center}
\epsfig{file=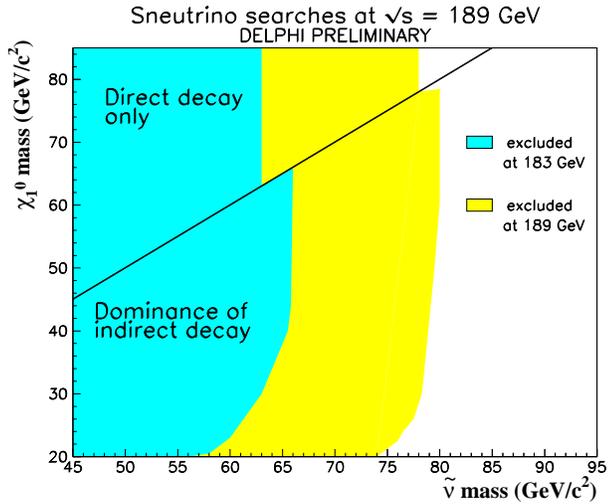,width=8.5cm} 
\caption[]{DELPHI limit on sneutrino pair production as allowed by \Rp\
with a dominant $\lambda_{133}$ coupling.}
\label{fig:snudouble}
\end{center}
\end{figure}

  All final states expected from pair production have been searched for.
An example of result for a dominant  $\lambda_{133}$ coupling is given
in Figure~\ref{fig:snudouble}. Using also the limit on \mkione, exclusion
limits can be derived for each of the three \Rp\ coupling classes. The
current limits~\cite{rplep} are presented in Table~\ref{tab:snu}. They
are valid for all values of the underlying parameters, except A$_0$ which
is set to 0.

\begin{table}[htb]
\begin{center}
\begin{tabular}{|c|c|c|c|}\hline
  \Rp\ coupling & experiment & limit on \msnu  & \snu\ flavour \\
  type          &            & (\Gevm)         & \\ \hline
  $\lambda$ couplings:     & DELPHI& 78 & any \snu\ flavour \\ 
  $\lambda^{'}$ couplings: & ALEPH & 56 & any \snu\ flavour\\ 
  $\lambda^{''}$ couplings:& ALEPH & 77 &  \snue\ only \\ \hline
\end{tabular}
\caption[]{LEP limits on \snu\ masses in the \Rp\ scheme,
from data up to 189~GeV.}
\label{tab:snu}
\end{center}
\end{table}

\begin{figure}[htb]
\begin{center}
\begin{tabular}{c|c}
\epsfig{file=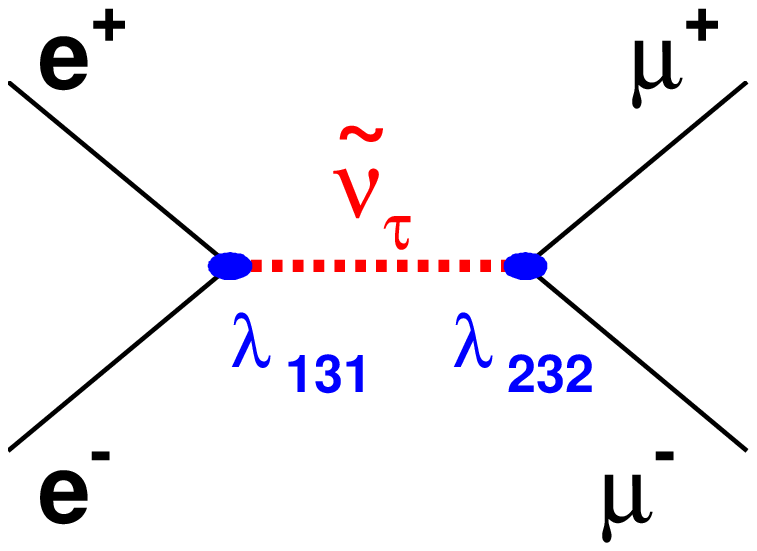,height=3cm}   &
\epsfig{file=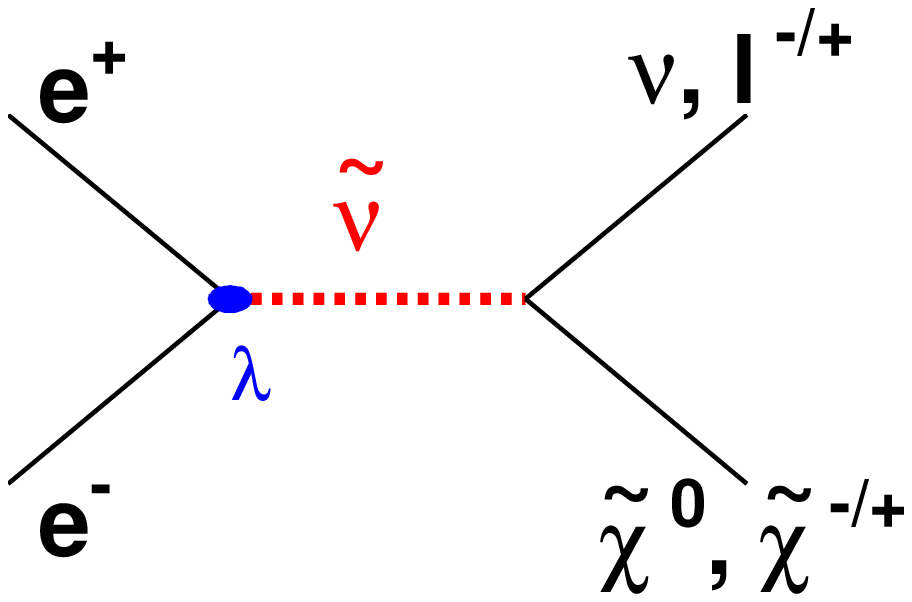,height=3cm} \\
Direct decays & Indirect decays \\
\end{tabular}
\caption[]{Single production of sneutrinos at LEP as allowed by \Rp\ .}
\label{fig:diagsnu}
\end{center}
\end{figure}

  Single production of \snum\ or \snut\ sneutrinos is also possible at
LEP, as illustrated in Figure~\ref{fig:diagsnu}.
Sneutrino direct decays would lead to effects observable as deviations 
wrt SM expectations while indirect decays would manifest as specific
final states that require dedicated searches. In both cases, 
constraints on sneutrino
masses have been derived as a function of the \Rp\ couplings.
Currently, masses between 100 to 200~\Gevm\ are probed and limits
on couplings are of the order of a few 10$^{-2}$~\cite{rpsnu}.

\subsubsection{Charged sleptons and squarks}

  Charged sleptons and light squarks are searched for at LEP. The coverage of
the final states expected in case of \Rp\ (from direct and indirect decays and
for any type of couplings) is not yet as complete  as for the other
sparticles. The results achieved so far~\cite{rplep} are summarized in
Table~\ref{tab:rpother}. As for other LEP results, they hold for
A$_0 = 0$, irrespective of the values of the other underlying parameters.

\begin{table}[htb]
\begin{center}
\begin{tabular}{|c|c|c|c|}\hline
  \Rp\ coupling & experiment & sparticle & mass limit \\
  type          &            &           & (\Gevm)    \\ \hline
  $\lambda$ couplings:     & ALEPH & \selr\ & 84 \\
  $\lambda$ couplings:     & ALEPH & \smur, \staur\ & 60 \\
  $\lambda^{'}$ couplings: & OPAL  & \stoone\ & 84 \\
  $\lambda^{''}$ couplings:& OPAL  & \stoone\ & 79 \\ \hline
\end{tabular}
\caption[]{LEP limits on charged slepton and light squark masses 
in the \Rp\ scheme, from data up to 189~GeV. The \stoone\ limits are valid
for any value of the stop mixing angle.}
\label{tab:rpother}
\end{center}
\end{table}

  Finally, there are also constraints on heavy squarks and gluinos if \Rpar\
is violated. LEP experiments reinterpret their results about leptoquark
searches~\cite{rpsq} while dedicated searches are performed at HERA and
Tevatron for particular \Rp\ couplings. HERA experiments have access mostly to
$\lambda^{'}_{1j1}$ couplings and derive constraints on heavy squark 
masses as a fuction of $\lambda^{'}_{1j1}$, varying the MSSM underlying 
parameters. Currently, masses around 240~\Gevm are excluded for 
$\lambda^{'}_{1j1} \sim 0.3$~\cite{rphera}. 
Using multilepton events in run~I data,
Tevatron experiments have probed $\lambda^{'}_{1jk}$ and $\lambda_{121}$
couplings, and constrained gluino and heavy squark masses. The limits
achieved are around 250 (resp. 350)~\Gevm\ for the $\lambda^{'}_{1jk}$
(resp. $\lambda_{121}$) coupling analysis~\cite{rptevatron}.
These limits are valid for
A$_0=0$, $\mu <0$ and \tb = 2. Most of the other choices 
(in particular, higher values of \tb\ or $\mu >0$) would lead to lower 
limits, due to a loss in sensitivity 
(resulting from reduced branching fractions into leptons, 
softer leptons...)~\cite{rptevatron}.

\subsection{Gauge mediated supersymmetry breaking}

  The third theoretical framework assumes \Rpar\ conservation and soft 
supersymmetry breaking mediated by gauge interactions. Such models~\cite{gg}
usually need six basic parameters: 
the supersymmetry breaking scale, $\sqrt{\rm{F}}$,
the universal mass scale of supersymmetric particles, $\Lambda$,
the messenger mass scale, $\rm{M_s}$, 
the number of messenger generations, $\rm{n_s}$,
the  Higgs mixing parameter, $\mu$,
and the ratio of the two Higgs doublet vacuum expectation values, \tb.
The breaking scale is expected to be much lower than in gravity-mediated 
models, down to about 10$^4$~GeV.

  As far as phenomenology at low energy is concerned, \Rpar\ conservation
implies as usual sparticle pair production and a stable LSP. As a
consequence of gauge-mediated breaking, the LSP is the gravitino, \sgrav,
whose mass depends on $\sqrt{\rm{F}}$ and thus is expected to be very
small, in the range [10$^{-6}$~eV, 1~keV]. The next lightest sparticle
(NSLP) is either \kione\ or a charged slepton (\stau\ or three degenerated
\sle). The NSLP lifetime is governed by the \sgrav\ mass and hence can be non
negligible, giving rise to specific topologies, some being experimental
challenges.

\subsubsection{\kione\ NLSP}
  
  The main decay of a \kione\ NLSP would be in $\gamma$\sgrav. Thus \kione\
searches rely on final states with photons, either single-photon or 
diphoton events. Such final states provide clean experimental signatures and
have been used since long to chase new physics, whatever the underlying 
theoretical framework. Results are usually expressed as model-independent
upper limits on the cross-section times branching fraction product,
as a function of the mass of the unknown particle decaying to a photon plus
missing energy. These cross-section limits are then compared with
predictions from various models~\cite{gmsbphot,gmsbtev}, such as the
GMSB scenario explaining an ee$\gamma\gamma$ event reported by CDF some years
ago. To quote but one result, even if not strictly a GSMB model, cross-section
limits have been converted into a lower limit on the mass of a superlight
\sgrav, assuming all other sparticles to be above threshold. The current
limit from single-photon events at LEP2~\cite{gmsbphot} is 
10$^{-5}$~eV/$c^2$, similar to the result reached by CDF using monojet 
events~\cite{cdfgrav}.
 
  Topologies more specific to GMSB models have also been searched for. The
first example is given by \kione\ searches in case of  long neutralino
lifetimes. The decay photons would not be produced at the interaction point 
but could have large impact parameters. Searches for non-pointing single-photon
events~\cite{gmsbphot} cover that case.
Searches for sparticles heavier than \kione\ have also been performed,
so far for charginos and sleptons~\cite{gmsbother} only. Compared with the
topologies expected in the gravity-mediated scheme, the final states
are identical except for additional photons that can be detected if the NLSP 
lifetime is not too long. As photons help to better discriminate against 
background, the exclusion limits for negligible or moderate NSLP lifetimes
are usually tigther than those in the gravity-mediated framework.

\subsubsection{\stau\ NLSP}

\begin{figure}[htb]
\begin{center}
\epsfig{file=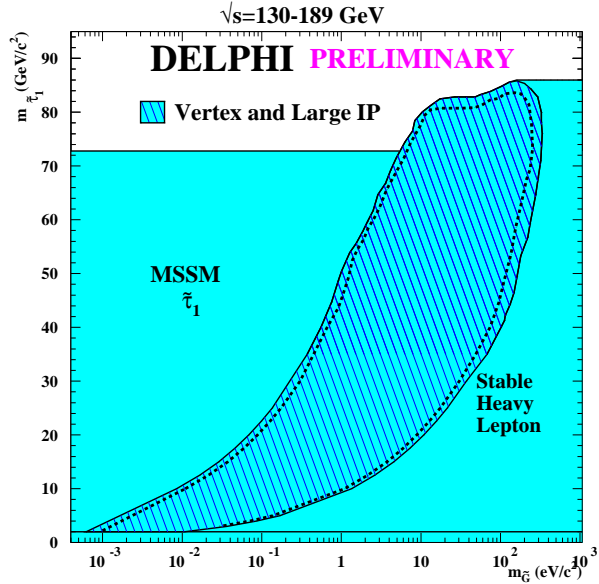,height=8.5cm} 
\caption{DELPHI lower limit on the \stau\ mass as a function of the \sgrav\
mass in GMSB models, from data up to 189~GeV.}
\label{fig:staus}
\end{center}
\end{figure}

  The main decay of a \stau\ NLSP would be in $\tau$ \sgrav, thus giving 
the same experimental signature as in the gravity-mediated case only if the
\stau\ lifetime is negligible. In the contrary case, \stau\ decays would
lead to kinks, large impact parameters or decay vertices. Eventually, a 
\stau\ decaying outside the detector would appear as a stable charged 
particle. All these experimental signatures have been used in \stau\ 
searches~\cite{gmsbother}. An example of result is given in 
Figure~\ref{fig:staus} which illustrates the interplay of the different
signatures as a function of the \sgrav\ mass which defines the \stau\ 
mean decay length. Irrespective of the \sgrav\ mass, these results
exclude a \stau\ NLSP up to 73~\Gevm. The limit is 6~\Gevm\ higher if
the results are reinterpreted in a scenario with three degenerated
co-NLSP charged sleptons.

Searches for sparticles heavier than \stau\ have also been performed,
so far for charginos, neutralinos and sleptons~\cite{gmsbother} and only
for negligible NLSP lifetimes.

\subsubsection{Constraining underlying parameters}

  Even if not complete, the coverage of the final states expected from GMSB
models is at present sufficient to exclude large fractions of the parameter 
space in order to set constraints on the key parameters of the model, 
like the NLSP mass or $\Lambda$. A first attempt has been reported 
in~\cite{gmsbscan} for a minimal GMSB model.

\subsection{Other searches, recent developments}

 To conclude about supersymmetry, it is worth mentioning a few complementary
results from LEP, either recent ones or results aside the main stream of
searches described in the previous sections.

  There is an ongoing effort to combine the four experiment results (on 
\charg, \selr, \stau\ searches, using also Higgs searches and the $\gz$ 
constraint) to derive an absolute limit on the \kione\ mass in the 
mSUGRA framework. 
Preliminary results~\cite{lepcsept} from data up to 189~GeV give a 
limit at 44~\Gevm\ if the difference between the \stau\ and \kione\ masses
is above 5~\Gevm\, irrespective of the values of the underlying parameters, 
except A$_0$ which is set to 0. Recently, dedicated searches have been 
performed to cover mass differences below 5~\Gevm\ and first results 
confirmed the validity of the above limit also in that case~\cite{lepcnov}. 
The next step should be to check the impact of A$_0$, which is expected 
to be large.

  About \Rp\, the superpotential quoted in equation~\leqn{eq:rp} is not 
the most complete \Rpar\ violating potential. Extra bilinear terms of the form
$\epsilon_i \rm{L}_i \rm{H}_2$, where H$_2$ is the Higgs superfield
with positive hypercharge, are also possible candidates to generate
\Rp\ . First results have been reported recently on a search for 
stops decaying through a bilinear term 
into a b$\tau$ pair~\cite{rpbilin}.

  Finally, if \Rpar\ is conserved, the LSP being stable
restricts the experimental sensitivity to other sparticles to mass
differences between the sparticle and the LSP in excess of a few~\Gevm.
Searches have been conducted to explore nearly mass degeneracy cases for
the lightest chargino~\cite{degencharg} and more recently for the lightest
selectron~\cite{degenselr}. Note that sparticles degenerated with the 
LSP exist in restricted regions of the parameter space of the constrained 
MSSM as defined in section~\ref{sec:cmssm} but
they are mostly predicted in other supersymmetric scenarios, such as a 
constrained MSSM without gaugino mass unification at the GUT scale
or the recent anomaly mediated SUSY breaking models~\cite{gg}.

\section{Search for extra dimensions}

  It was pointed out recently~\cite{hierarchy,gg} that extra spatial
dimensions, which are present in any superstring theory, can also solve 
the hierarchy problem, independently of the underlying theoretical
framework. Indeed, if n extra compact spatial dimensions of radius R
exist, the quantum gravity scale in n+4 dimensions, M$_{\rm D}$, is
related to the Planck scale, M$_{\rm Pl}$, by 
 M$_{\rm Pl}^2 \sim {\rm R}^n {\rm {M}^{2+n}_{D}}$.
If R and n are such that M$_{\rm D}$ is of the order of the 
electroweak (EW) scale, the
hierarchy vanishes. The case n=1 is ruled out since it would imply
quantum gravity effects observable over solar system distances.

  At low energy, extra spatial dimensions are expected to manifest through
the production of gravitons, G, observable in both direct searches and
precise measurements. Searching for the associated production of a pair
($\gamma$, G) in single photon final states at LEP2 leads to constraints
on  M$_{\rm D}$ depending on n. As an example,  M$_{\rm D}$ has been found 
to be greater than 1.1~TeV, 0.7~TeV and 0.53~TeV for 2, 4 or 6 extra 
dimensions, respectively~\cite{extradim}. Gravitons would also be responsible
for deviations wrt the SM in precise measurements. Combining observables in
several final states at LEP, the ultra-violet cut-off of the underlying 
quantum gravity theory has been constrained to be larger than 0.8~TeV 
(resp. 1.1~TeV) if the interference between the SM and G exchange amplitudes
is negative (resp. positive)~\cite{extradim}.

\section{Higgs bosons}
  
   The phenomenology of Higgs bosons is little model-dependent 
which allows to cover several
theoretical frameworks with a limited number of searches.
Results encompass neutral Higgs bosons as expected in the SM, 
MSSM and beyond, as well as charged Higgs bosons in 
two Higgs doublet models.

\subsection{The Standard Model Higgs boson}

  In the mass range currently under study, \ie\ around 100~\Gevm, only LEP
would be sensitive to the SM Higgs boson. The main production process leads
to pairs of Higgs and Z bosons, with the Z boson on-shell. Due to the clean 
experimental environment, all Z final states are exploited. In addition,
the Higgs boson is expected to decay mainly into a 
${\rm b}{\bar {\rm b}}$ pair (the branching fraction is $\sim$ 82\% for 
a 100~~\Gevm\ Higgs boson) so that excellent b-tagging capabilities of the
LEP detectors help a lot in these searches.

\begin{table}[htb]
\begin{center}
\begin{tabular}{|c|c|c|c|c|c|} \hline
& bkg  & data & exp. limit & obs. limit &  1-CL$_b$        \\
&      &      & (~\Gevm)   & (~\Gevm)   &  at obs. lim. \\ \hline
ALEPH  & 44.4  & 53  & 95.9 & 92.9 & 4$\%$ \\
DELPHI & 172.7 & 187 & 94.6 & 94.1 & 20$\%$ \\
L3     & 91.1  & 94  & 94.8 & 95.3 & 64$\%$ \\
OPAL   & 35.4  & 41  & 94.9 & 91.0 & 4$\%$ \\ \hline
\end{tabular}
\caption{Rates and exclusion limits obtained at LEP in the SM Higgs
boson searches in 189~GeV data. Results observed in data 
are compared with expectations from background simulation. 
The last column gives the probability 
to have a less background-like result than that observed. }
\label{tab:higgssm}
\end{center}
\end{table}

\begin{figure}[htb]
\begin{center}
\epsfig{file=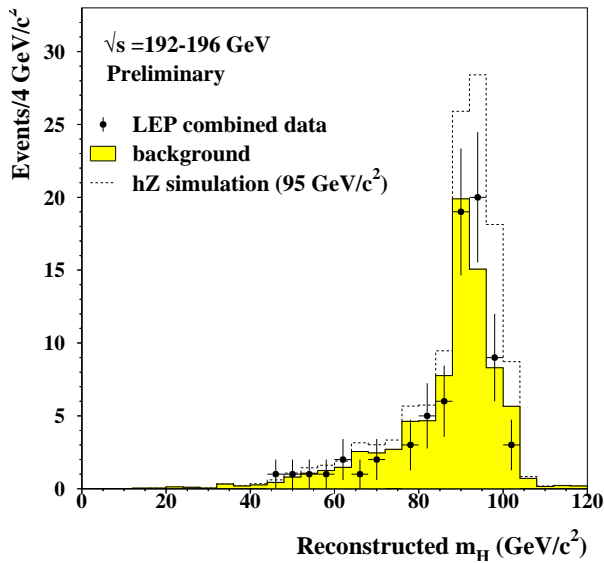,height=8cm}
\caption{Reconstructed Higgs boson mass spectrum in LEP data at 192 and
196~GeV. Data (dots) are compared with background simulation (full histogram) 
and with signal simulation in the hypothesis of a SM Higgs boson of 95~\Gevm
(dashed histogram). 74 events are selected in data while 81.1 are expected
from background.}
\label{fig:higgssm}
\end{center}
\end{figure}

\begin{table}[htt]
\begin{center}
\begin{tabular}{|c|c|c|c|c|c|} \hline
& $\L$       & bkg & data & exp. limit & obs. limit \\
& (pb$^{-1}$)&     &      & (~\Gevm)   & (~\Gevm) \\ \hline
ALEPH  & 98  & 32.3 & 27 & 99.9 & 98.8 \\
DELPHI & 84  & 15.4 & 15 & 97.0 & 97.3 \\
L3     & 109 & 42.2 & 38 & 97.3 & 98.7 \\
OPAL   & 85  & 21.0 & 23 & 97.3 & 95.4 \\ \hline
\end{tabular}
\caption{Rates and exclusion limits obtained at LEP in the SM Higgs
boson searches in data at 192 and 196~GeV. Results observed in data 
are compared with expectations from background simulation.}
\label{tab:higgssm99}
\end{center}
\end{table}

  The results obtained at 189~GeV by the four LEP experiments~\cite{higgs189} 
are detailed in Table~\ref{tab:higgssm} at the level of selections where data
are actually compared with simulation to test the background and 
signal+background hypotheses and derive exclusion limits or discovery 
significances. To achieve the highest sensitivity to the signal, 
these derivations rely on a test-statistic which, besides the rates, 
take also into account the pattern of the selected events
(the reconstructed Higgs boson mass, m$_{\rm H}$, or m$_{\rm H}$ and another
discriminant variable like the event b-quark content)~\cite{hwg}.
The more information in the comparison, the earlier the event 
selection procedure is stopped, as illustrated in Table~\ref{tab:higgssm} 
by the different selection levels in the four experiments. 
There is an excess in data in three experiments, which is partly signal-like 
in two of them, as revealed by the difference of a few~\Gevm\ between the 
observed and expected exclusion limits. After investigation, part of the 
excess was attributed to a systematic bias. When combining 
the four LEP experiments, the exclusion limit is 95.2~\Gevm\ for an 
expectation of 97.2~\Gevm~\cite{hwg189}. 

  A preliminary update at 196~GeV was reported at the 
conference~\cite{higgs196}, as shown in Table~\ref{tab:higgssm99}. 
The reconstructed Higgs boson mass distribution after tighter selections 
is given in Figure~\ref{fig:higgssm}. The excess seen at 189~GeV has not 
been confirmed at higher energies and data agree with expectations. 
After the conference, these results were combined, giving an exclusion limit 
of 102.6~\Gevm, for an expected limit of 102.3~\Gevm~\cite{hwg196}.
At the end of the '99 run, which reached \rs=202~GeV, experiments reported 
preliminary limits up to 106~\Gevm, in good agreement with the expected 
ones~\cite{higgs202}. Prospects for the last run of LEP at \rs\ up to
206~GeV are 114~\Gevm\ for the 95\% exclusion or 3$\sigma$ discovery
potentials and 111~\Gevm\ for the 5$\sigma$ discovery sensitivity, when
the four experiments are combined~\cite{chamonix00}. Higher masses
up to 180~\Gevm\ should then be accessible in the future
high luminosity run at the Tevatron~\cite{run2wkp}.

\subsection{MSSM neutral Higgs bosons}
  Most results about neutral Higgs bosons in the MSSM come also from LEP, 
which is sensitive to the two lightest bosons, h and A. There are two 
production processes, \eea hZ, like in the SM case, and \eea hA. 
The two processes are complementary in the parameter space. In the mass 
range between 80 and 110~\Gevm, the main decay mode of both bosons is 
again in ${\rm b}{\bar {\rm b}}$ in most of the parameter space, 
with branching fractions greater than in the SM ($\sim$ 91\%). The dominant
hZ final states are as in the SM case, while hA is expected to give mostly 
${\rm b}{\bar {\rm b}}{\rm b}{\bar {\rm b}}$ and 
$\tau^+\tau^-{\rm b}{\bar {\rm b}}$ final states. Here again, b-tagging plays
a crucial r\^ole.
  
  The theoretical framework of these searches is the MSSM with \Rpar\ 
conservation and soft breaking terms unified at the EW scale. In the MSSM 
the Higgs boson masses are connected to each other, so that at tree
level, there are only two free parameters: \tb\
and one Higgs boson masses, or, alternatively, two Higgs boson 
masses, eg  \mA\ and \mh. The properties of the MSSM Higgs bosons, and in 
particular the mass relationships, are modified by radiative corrections
which introduce five additional parameters: 
the mass of the top quark, 
the Higgs mixing parameter, $\mu$, 
the common sfermion mass term at the EW scale, M$_S$, 
the common SU(2) gaugino mass 
term~\footnote{The U(1) gaugino mass term at the EW scale, 
M$_1$, is related to M$_2$ through the GUT relation
M$_1 = (5/3) \rm{\tan}^2\theta_w \rm{M_2}$, while the SU(3) 
gaugino mass term, M$_3$, is set via the gluino mass, which is taken equal to
M$_S$.} at the EW scale, M$_2$,
and the common squark tri-linear coupling at the EW scale, A. 
The interpretation of the experimental results depend on the
values assumed for these parameters.

\subsubsection{Benchmark hypotheses}

  Using leading order two-loop calculations of the radiative 
corrections~\cite{onelooprc}, benchmark values have been defined for the
parameters beyond tree-level~\cite{lep2wk_pres}: 
175~\Gevm\ for the top mass, 1~TeV/c$^2$ for M$_S$ and 1.6~TeV/c$^2$ 
for M$_2$. 
Two benchmark 
scenarios~\cite{lep2wk_pres} have been defined for the parameters 
A and $\mu$, which determine the mixing in the stop sector:
no mixing (A$ = 0$, $\mu = -100~$~GeV) and
maximal mixing (A$ = \sqrt{6} \rm{M}_S$, $\mu = -100~$~GeV).
The no mixing hypothesis leads to minimal radiative corrections to \mh\
while the maximal mixing induces the largest corrections. This defines
the usual framework for the interpretation of the MSSM neutral Higgs
boson searches.

\begin{figure}[htb]
\begin{center}
\epsfig{file=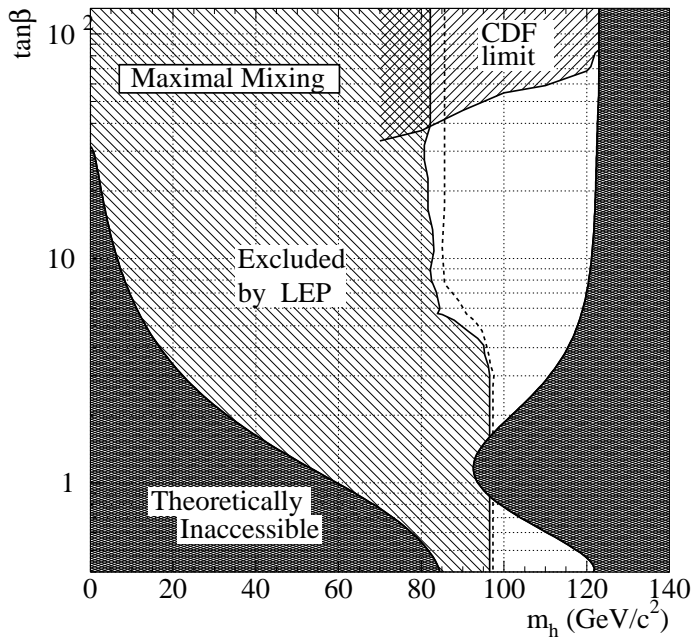,height=10cm}
\caption{Regions in the (\mh, \tb) plane excluded by the MSSM Higgs boson 
searches at LEP in data up to 189~GeV, and at CDF in run I data. 
The regions not allowed by the MSSM for a
top mass of 175~\Gevm\, a SUSY scale of 1~TeV/c$^2$
and maximal mixing in the stop sector are also indicated.
The dotted curve is the LEP expected limit.}
\label{fig:higgsmssm}
\end{center}
\end{figure}

 Results obtained at LEP at 189~GeV in the hA channel are summarized
in Table~\ref{tab:higgsmssm}. These results, together with the hZ results 
reinterpreted in the MSSM framework, allow to set constraints on
\mh, \mA\ and \tb. As an example, Figure~\ref{fig:higgsmssm} represents
the region of the (\tb, \mh) plane excluded by the combination of the
LEP results up to 189~GeV, in the less favourable
case of the maximal mixing~\cite{hwg189}. Whatever the mixing hypothesis, 
these combined results exclude Higgs bosons up to 80.7~\Gevm\ for  \mh\
and 80.9~\Gevm\ for \mA\, for \tb\ greater than 0.4. At \tb$\sim 1$, the
experimental lower limit on  \mh\ is above the theoretical upper bound
on  \mh\, so that \tb\ is excluded between 0.9 and 1.6 (0.6 and 2.6)
in the maximal mixing (no mixing) hypothesis. The expected limits  
on both masses are 5~\Gevm\ higher while the expected excluded ranges 
in \tb\ agree with the observed ones. 

\begin{table}[htb]
\begin{center}
\begin{tabular}{|c|c|c|c|c|c|c|} \hline
& bkg & data & \multicolumn{2}{c|}{\mh\ limits (~\Gevm)} 
             & \multicolumn{2}{c|}{\mA\ limits (~\Gevm)} \\
&     &      &  observed & expected  &  observed & expected \\ \hline
ALEPH  & 7.5  & 10  & 82.5 & 83.1 & 83.1 & 83.6 \\
DELPHI & 22.6 & 24  & 82.1 & 81.1 & 83.1 & 82.2 \\
L3     & 140.6 & 153& 76.0 & 78.0 & 76.0 & 79.0 \\
OPAL   & 12.9 & 15 & 74.8  & 76.4 & 76.5 & 78.2 \\ \hline
\end{tabular}
\caption{Rates and exclusion limits obtained at LEP in the MSSM hA Higgs
boson searches in 189~GeV data. Results observed in data 
are compared with expectations from background simulation.}
\label{tab:higgsmssm}
\end{center}
\end{table}
 
   It must be noted that, contrary to the limits on masses, the limit on 
\tb\ is very sensitive to the values of the underlying parameters, which have 
a large impact on the theoretical upper bound on \mh. As an example, this
upper bound increases with increasing top quark masses and no limit
on \tb\ is obtained if the top mass is moved by two standard deviations.
The pure MSSM parameters or the order of the radiative
correction calculations have also a non negligible effect. So, the excluded 
ranges in  \tb\ cannot be taken as an absolute result, even if the maximal
mixing hypothesis is a pessimistic scenario.

  Also displayed in Figure~\ref{fig:higgsmssm} is the recent CDF 
result~\cite{higgscdf} at large \tb. In this region, the production 
cross-section is large enough to make Tevatron sensitive to the production
process \pp ${\rm b}{\bar {\rm b}}$ h,H,A where one Higgs boson is emited 
off a bottom quark, leading to four b-tagged jets in the final state.

  Results from LEP experiments with data up to 196~GeV were also reported at 
the conference~\cite{higgs196}, as shown in Table~\ref{tab:higgsmssm99}. 
These results were combined later on, giving exclusion limits of 
84.3~\Gevm\ on \mh\
and 84.5~\Gevm\ on \mA, independent of the mixing hypothesis, and
excluded ranges in \tb\ between 0.8 and 1.9 (0.5 and 3.2) in the maximal 
mixing (no mixing) hypothesis~\cite{hwg196}. At the end of the '99 data 
taking, LEP experiments reported mass limits up to 90~\Gevm\ and excluded
ranges in \tb\ somewhat larger than the combined result at 
196~GeV~\cite{higgs202}. The excess in data observed by OPAL in the
hA channel at 196~GeV was not confirmed at higher energies nor by the 
other experiments.

\begin{table}[htb]
\begin{center}
\begin{tabular}{|c|c|c|c|c|c|} \hline
& $\L$       & bkg & data & \multicolumn{2}{c|}{\mh\ limits (~\Gevm)} \\
& (pb$^{-1}$)&     &      & observed & expected \\ \hline
ALEPH   & 98  & 4.8 &  1   & 85.2 & 86.1 \\ 
OPAL    & 85  & 6.8 & 14   & 74.3 & 79.1 \\ \hline 
\end{tabular}
\caption{Rates and exclusion limits obtained at LEP in the MSSM hA Higgs
boson searches in 192 and 196~GeV data. Results observed in data 
are compared with expectations from background simulation.}
\label{tab:higgsmssm99}
\end{center}
\end{table}

\subsubsection{General scans}

  More general interpretations of the LEP Higgs boson searches have been
performed, scanning over the MSSM underlying parameters. The parameter
space is however usually restricted by imposing additional constraints, \eg\
the experimental results on supersymmetric particles or the $\gz$ constraint.
It was shown that the benchmark limits hold in more than 99.99\% of
the parameter sets~\cite{fullsc183} and that the mass limits from
general scans are a few~\Gevm\ weaker than the benchmark ones~\cite{fullsc189}.
As general scans usually vary also the top mass quark within two standard
deviations, the limit on \tb\ vanishes.

\subsubsection{Recent developments}

  Recent theoretical work led to two-loop calculations of the 
radiative corrections at the next-to-leading order and 
to a  redefinition of the benchmark values for the
underlying parameters~\cite{twolooprc}. In particular, the theoretical upper
bound on \mh\ was found to be underestimated by $\sim$7~\Gevm\ in the 
maximal mixing scenario previously used. The new scenario  now proposed 
(called \mh$^{max}$ scenario) should lead to more realistic bounds on
\tb. Another new scenario (called large $\mu$ scenario) with the h boson
within kinematical reach at LEP but with vanishing branching fraction 
into ${\rm b}{\bar {\rm b}}$ was proposed to check the sensitivity of LEP
to Higgs bosons with non-dominant b decays. Future LEP results will be derived
in these new benchmark schemes.

\subsection{Neutral Higgs bosons beyond MSSM}

  Searches for neutral Higgs bosons as expected beyond the MSSM have also
been performed, mostly at LEP. Three lines of searches have been followed.
First, the existing LEP analyses on MSSM h and A bosons have been used,
either as such or with some modifications (\eg\ relaxing the b-tagging 
requirements) to cover the final states expected in more general models.
Thus, a first study showed that LEP is sensitive to neutral Higgs
bosons of two Higgs doublet models (2HDM), even in a scenario with 
dominant decays into $\rm{c}\bar{\rm c}$ or in a model with CP 
violation~\cite{2hdmd}. Recently, the 2HDM parameter space 
(with CP conservation) has been explored in a detailed scan~\cite{2hdmo}.
Finally, for the first time, a non minimal supersymmetric model 
containing one gauge-singlet Higgs field in addition to the MSSM has 
also been investigated~\cite{nmssm}.

  As a second research line, the case of a Higgs boson h decaying 
invisibly has been studied. Dedicated searches in the hZ channel
translate into upper limits on the production cross-section times 
branching ratio, which are compared with expectations from specific
models~\cite{invis}. As an example, assuming a SM production rate
and a 100\% branching ratio into invisible products, a lower limit on
\mh\ at 95.4~\Gevm\ is obtained.

  The third topic deals with a Higgs boson h with anomalous couplings
to photons. From dedicated searches in the hZ, h$\gamma$ and hA channels,
general constraints are set on the production cross-section times branching
ratio or directly on the anomalous couplings. They are again compared
with expectations from specific models~\cite{higgsphot,higgs196}. 
As an example, 
assuming a SM production rate and a fermiophobic Higgs boson, a lower 
limit on \mh\ at 97.5~\Gevm\ is achieved.

\subsection{Charged Higgs bosons}

  Recent results on charged Higgs bosons, \hpm, have been reported by the LEP
experiments. The framework of these searches is the general 2HDM scheme
with as sole free parameters the \hpm\ mass and its leptonic decay branching 
fraction, assuming that the hadronic (into cs) and leptonic 
(into $\tau \nu_{\nu}$) decays saturate the width of the particle, 
which is the case in the mass range below $\mw$ that is presently tested.

\begin{table}[htb]
\begin{center}
\begin{tabular}{|c|c|c|c|c|c|} \hline
& bkg & data & expected limit & observed limit \\
&     &      & (\Gevm) & (\Gevm) \\ \hline
ALEPH & 333.5  & 302 & 69.5 & 65.5 \\
DELPHI & 213.0  & 215 & 66.5 & 66.9 \\
L3 & 523.5  & 499 & 71.2 & 67.5 \\
OPAL & 241.1  & 252 & 68.5 & 68.7 \\ \hline
\end{tabular}
\caption{Rates and exclusion limits obtained at LEP in the charged Higgs
boson searches in 189~GeV data. Results observed in data 
are compared with expectations from background simulation.}
\label{tab:hphm}
\end{center}
\end{table}

  The results obtained by the LEP experiments at 189~GeV~\cite{hphm189}
are shown in Table~\ref{tab:hphm}. Once combined, these results exclude 
an \hpm\ boson up to 77.3~\Gevm\ whatever its leptonic decay branching ratio,
while the expected limit is 74.9~\Gevm~\cite{hwg189}. Results from data 
up to 196~GeV were combined after the conference, with no improvement in 
the mass limit independent of the branching ratio, due to the large 
background from WW pairs which is penalizing for the analyses in the 
hadronic mode. On the other hand, \hpm\ bosons with pure (50\%) leptonic 
decays have been excluded up to 84.9~(78.4)~\Gevm\ by the same 
results~\cite{hwg196}. Individual limits reported at the end of the 
'99 data taking were below the combined results at 196~GeV~\cite{higgs202}.

  Higher masses were tested at the Tevatron runI, searching for
\hpm\ in decays of pair-produced top quarks. The mass limits are \tb\
dependent and restricted to those values of \tb\ for which the top quark
branching fraction into \hpl b is large enough. In the most favourable 
case (\tb=150) masses up to 153~\Gevm\ have been excluded~\cite{hphmtev}.

\section{Conclusions}

  New particle searches cover an impressive variety of topics and 
topologies. The way results are interpreted has undergone substantial
changes during the past few years. To get higher sensitivity to
the researched signals, different channels and/or experiments are
combined and more information is put in the statistical analysis
of the results, like in the Higgs boson searches. There is also an
effort to go to more model-independent results by relaxing theoretical
assumptions, scanning parameter values or testing more general models.
Supersymmetric particle searches are examples of that kind.

  To give but a few results, in gravity-mediated SUSY breaking models, 
the lightest neutralino has been excluded up to 32~\Gevm\ whether \Rpar\ is 
conserved or not, while the lightest chargino has been excluded up to
68~(94)~\Gevm\ if \Rpar\ is conserved (broken). A SM Higgs boson has
been excluded up to 106~\Gevm, MSSM neutral Higgs bosons up to 90~\Gevm
and 2HDM charged Higgs bosons up to 77~\Gevm.

\bigskip
I am grateful to all my colleagues who provided me with information
when preparing this talk.
I would like to thank  M.Besancon, R.Nikolaidou, E.Perez and D.Treille for
very helpful discussions.

\def\Discussion{
\setlength{\parskip}{0.3cm}\setlength{\parindent}{0.0cm}
     \bigskip\bigskip      {\Large {\bf Discussion}} \bigskip}
\def\speaker#1{{\bf #1:}\ }

\Discussion

\speaker{Lee Roberts (Boston University)}
Can you comment again on the limits on tan$\beta$ from 
the Higgs searches?

\speaker{Ruhlmann-Kleider}  
The limits on tan$\beta$ have been derived with two-loop leading
order calculations
of the radiative corrections and for specific values of the underlying 
parameters so they cannot be regarded as absolute, even if the maximal 
mixing scenario represents a difficult case. Going to next-to-leading
order two-loop  
calculations, varying the underlying parameters or increasing the top 
quark mass will make the excluded range in tan$\beta$ decrease if not 
vanish.

\end{document}

%% file: LP99macros.tex
\def\beq{\begin{equation}}
\def\eeq#1{\label{#1}\end{equation}}
\def\eeqn{\end{equation}}

\def\beqa{\begin{eqnarray}}
\def\eeqa#1{\label{#1}\end{eqnarray}}
\def\eeqan{\end{eqnarray}}

\def\leqn#1{(\ref{#1})}

\def\Journal#1#2#3#4{{#1} {\bf #2}, #3 (#4)}

\def\PLB{Phys. Lett.  B}
\def\PRL{Phys. Rev. Lett.}
\def\PRD{Phys. Rev. D}

\let\bar=\overbar

\def\ie{{\it i.e.}}
\def\eg{{\it e.g.}}

\def\L{{\cal L}}

\def\O{{\cal O}}
\def\W{{\cal W}}

\def\Dslash{\not{\hbox{\kern-4pt $D$}}}
\def\dslash{\not{\hbox{\kern-2pt $\del$}}}

\def\gz{\Gamma_Z}
\def\mw{m_W}

\def\msb{{\bar{\ssstyle M \kern -1pt S}}}

